\documentclass[authorversion,sigconf,nonacm]{acmart}

\AtBeginDocument{%
  \providecommand\BibTeX{{%
    \normalfont B\kern-0.5em{\scshape i\kern-0.25em b}\kern-0.8em\TeX}}}

\copyrightyear{2022}
\acmYear{2022}
\setcopyright{acmcopyright}
\acmConference[CHI '22]{CHI Conference on Human Factors in Computing Systems}{April 29-May 5, 2022}{New Orleans, LA, USA} \acmBooktitle{CHI Conference on Human Factors in Computing Systems (CHI '22), April 29-May 5, 2022, New Orleans, LA, USA}
\acmPrice{15.00}
\acmDOI{10.1145/3491102.3501940}
\acmISBN{978-1-4503-9157-3/22/04}

\usepackage{_macros}

\usepackage{caption}
\usepackage{subcaption}
\usepackage{array}

\newcommand{\name}{VibEmoji}

\begin{document}

\title{\name: Exploring User-authoring Multi-modal Emoticons in Social Communication}


\author{Pengcheng An}
\authornote{These authors contributed equally.}
\orcid{0000-0002-7705-2031}
\affiliation{%
  \institution{School of Design, SUSTech}
  \city{Shenzhen}
  \country{China}
}
\affiliation{%
  \institution{University of Waterloo}
  \city{Waterloo}
  \state{Ontario}
  \country{Canada}
}
\email{anpengcheng88@gmail.com}

\author{Ziqi Zhou}
\authornotemark[1]
\author{Qing Liu}
\authornotemark[1]
\affiliation{%
  \institution{University of Waterloo}
  \city{Waterloo}
  \state{Ontario}
  \country{Canada}
}
\email{z229zhou@edu.uwaterloo.ca}
\email{qing.liu@uwaterloo.ca}

\author{Yifei Yin}
\affiliation{%
  \institution{University of Toronto Scarborough}
  \city{Scarborough}
  \state{Ontario}
  \country{Canada}
}
\email{yifei.yin@mail.utoronto.ca}

\author{Linghao Du}
\author{Da-Yuan Huang}
\affiliation{%
  \institution{Human-Machine Interaction Lab, Huawei Canada}
  \city{Markham}
  \state{Ontario}
  \country{Canada}
}
\email{linghao.du@huawei.com} \email{dayuan.huang@huawei.com}

\author{Jian Zhao}
\authornote{Corresponding author.}
\orcid{0000-0001-5008-4319}
\affiliation{%
  \institution{University of Waterloo}
  \city{Waterloo}
  \state{Ontario}
  \country{Canada}
}
\email{jianzhao@uwaterloo.ca}

\renewcommand{\shortauthors}{P. An, Z. Zhou, Q. Liu, Y. Yin, L. Du, D. Huang \& J. Zhao}

\begin{abstract}
Emoticons are indispensable in online communications. 
With users' growing needs for more customized and expressive emoticons, recent messaging applications begin to support (limited) multi-modal emoticons: \eg, enhancing emoticons with animations or vibrotactile feedback. However, little empirical knowledge has been accumulated concerning how people create, share and experience multi-modal emoticons in everyday communication, and how to better support them through design. 
To tackle this, we developed \name{}, a user-authoring multi-modal emoticon interface for mobile messaging. Extending existing designs, \name{} grants users greater flexibility to combine various emoticons, vibrations, and animations on-the-fly, and offers non-aggressive recommendations based on these components’ emotional relevance. 
Using \name{} as a probe, we conducted a four-week field study with 20 participants, to gain new understandings from in-the-wild usage and experience, and extract implications for design. We thereby contribute to both a novel system and various insights for supporting users' creation and communication of multi-modal emoticons.
\end{abstract}



\begin{teaserfigure}
  \includegraphics[width=\textwidth]{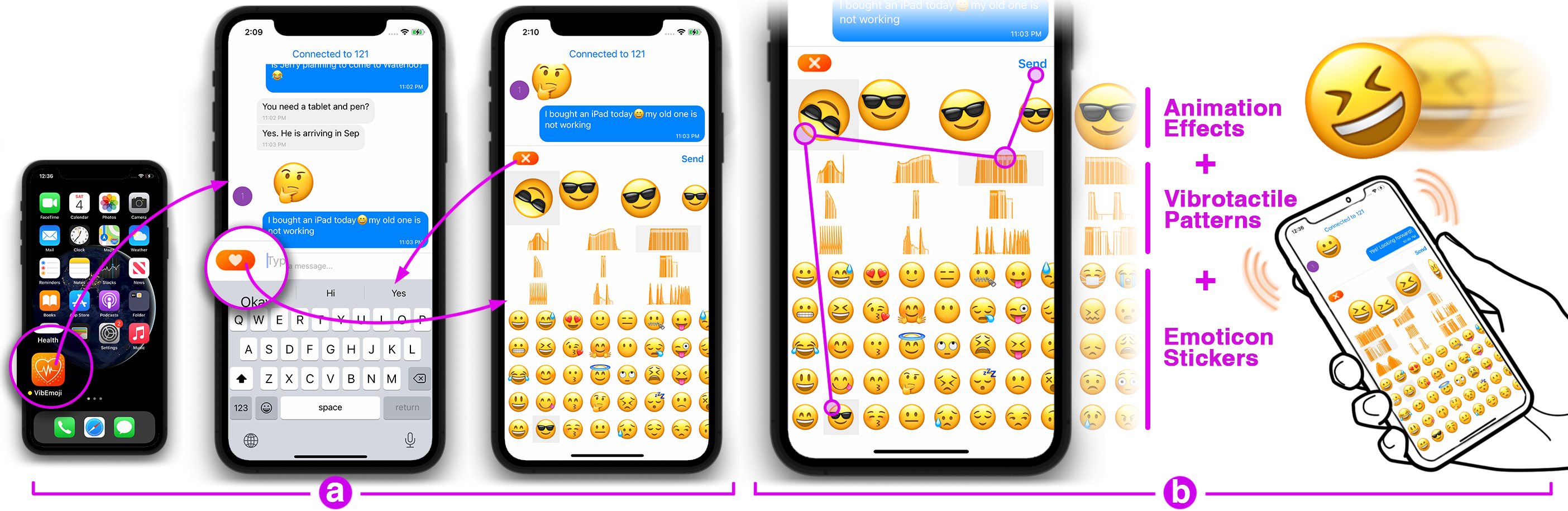}
  \vspace{-7mm}
  \caption{(a) \name{} enables users to seamlessly create and send multi-modal emoticons during messaging; (b) users could freely combine various stickers, vibrations, and animations to convey rich and nuanced feelings on the fly, in response to the unfolding conversation.}
  \Description{}
  \label{fig:teaser}
\end{teaserfigure}


\maketitle

\section{Introduction}

The use of emoticons has become ubiquitous, cross-cultural, and increasingly essential in online social communications. As a universal pictogram or visual language, emoticons facilitate communications within and across different linguistic and cultural backgrounds \cite{danesi2016semiotics}. Moreover, they enable people to express themselves concretely, conveying emotions, feelings, or non-verbal reactions that cannot be easily articulated by words. As a result, over 20\% of tweets now include an emoji \cite{tweetemojis}, and five billion emojis are sent everyday on Facebook Messenger \cite{facebookemojis}.

Along with the fast-growing popularity of emoticons, new emoticons still continue being demanded by users for various communication needs \cite{emojiRequest}. 
Meanwhile, the classic, static emoticons seem to no longer fully suffice people's various needs for expression. Users continue seeking further enriched, more expressive, and customized ways to communicate via emoticons. This trend could be well demonstrated by the recent emergence of \textit{multi-modal emoticons} across major social applications: \ie, emoticons augmented by multi-modal effects such as animations or vibrotactile feedback. For example, Apple users could customize the appearance of emoticons (Memoji \cite{applememoji}) and record animations for stickers (Animoji \cite{appleanimoji}). Google Gboard \cite{googlegboard} allows users' mashups of two emojis. Tencent WeChat \cite{tencentwechat} and Huawei MeeTime \cite{huaweimeetime} enable both animation and haptic effects of specific emoticons. The rapid emergence of multi-modal emoticons reflects users' desires for a more customized, multi-modal experience in social interaction.

Despite the burgeoning of multi-modal emoticons in consumer products, the HCI research field has accumulated little knowledge about how people create, share and experience multi-modal emoticons in daily communications and how to better support them by design. The exploration presented in this paper sets out to tackle this under-explored opportunity. 
Namely, we build and evaluate \name{}, a user-authoring multi-modal emoticon system that supports users to effortlessly author, and communicate via multi-modal emoticons during everyday messaging (\autoref{fig:teaser}). 
We employ \name{} as both a design to study about and a technology probe to study with, generating new empirical knowledge about the usage and user experiences of multi-modal emoticons, and extract relevant implications for future research and design in this domain. 

Given these explorative research purposes, \name{} is designed with several features that extend beyond, or differentiate from, existing design cases. First, in current systems (\eg, MeeTime, Telegram, and iMessage), the emoticon sticker, animation effect, and vibrotactile feedback are often fixed combinations; users cannot re-combine or appropriate these elements to create new meanings.

By contrast, in \name{}, users are allowed to freely select from and combine three rich sets of elements: pictograph emoticons, animation effects, and vibrotactile patterns (Figure \ref{fig:teaser}). This is to better understand users' creative usage in social communications \cite{Wiseman2018_repurposingemoji}, and support them to create more nuanced, personalized expressions in a conversation. 

Second, emoticon recommendation in current systems often tends to automate users' selection (\eg, typing ``happy'' and \wsicon{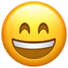} will be suggested), which could result in irrelevant suggestions due to misinterpreting the user's intent. Differently, \name{} explores a non-aggressive recommendation strategy that eases users' authoring without limiting their options. Each time when the user selects the first element (\eg, an emoticon), \name{} predicts the most likely elements to be selected in the remaining two categories (\eg, vibrations and animations), and updates the display order to prioritize these elements on the interface. The prediction is based on two sources of information: (1) emotional relevance between the selectable elements (base on the valence-arousal model \cite{mehrabian1980basic,russell1999core}), and (2) frequency of combinations authored by the user.

Third, current user-customization of multi-modal emoticons is often done in advance, separated from the moment of use. For instance, users need to pre-define the customizable parameters (\eg, Memoji), or pre-record an animation (\eg, Animoji), so that they could use it later. Differently, we design \name{} with the intention to support user-authoring on the fly, during a conversation, so that we could explore how users' creation of multi-modal emoticons can be improvised, based on their real-time feeling and context. 

To empower the design and implementation of \name{}, two questionnaire surveys have been conducted to gather perceived emotional properties of 15 animation effects (52 respondents) and 60 vibrotactile patterns (52 respondents). This is because adequate datasets remain to be established for vibrations and animations, besides the emotional properties of emoticons collected in \cite{rodrigues2018lisbon}.  
Within \name{}, we developed a mobile application and a back-end system, which were deployed in a field evaluation with 20 participants in 10 pairs, over the period of four weeks. Each pair of participants is friends or partners who have pre-existing routines of mobile messaging. A mixed-method approach was adopted to gather both quantitative and qualitative data to uncover empirical knowledge about how the participants created, used, and experienced multi-modal emoticons in their daily communications. Based on our exploration, we generalize a set of design implications of user-authoring, multi-modal emoticon systems, to inform future design and research.

This work thereby has twofold contributions: (1) a novel mobile system that supports users' authoring and usage of multi-modal emoticons in online communication;
(2) empirical knowledge about how people create, share and experience multi-modal emoticons in daily communication, and relevant implications for future designers and researchers.

\section{Background} \label{sec:background}

\subsection{Emoticons and Affective Communication}

While the most famous early example of emoticons,\space\space\textbf{:-)}\space\space, has been proposed by Fahman \cite{ScottFahlman} in his email to colleagues in 1982, the earliest use of emoticons in computing systems is dated back to the 1970s (the PLATO system \cite{plato}). 
While the term \textit{emoticon} and \textit{emoji} are often used interchangeably (as this paper also does), the name emoji is originally rooted in the Japanese mobile market (firstly supported in a product by J-Phone in 1997), which literally translates as ``facial 
letters/characters.''
As a universal, powerful pictographic language, emoticons/emojis have been widely used to concretely and conveniently convey emotional feelings that would otherwise take more words to articulate. 
For this reason, a great number of related studies analyzed the usage of emoticons among relatively large-scale samples to understand people's emotions in various contexts: \eg, to detect software developers' emotions in communication \cite{Chen2021_emotion}, to collect students' emotional states in learning \cite{Zhang2017_affectivestates}, or to understand people's political attitudes \cite{Hagen2019_emojiuse}, to name but a few.

Besides analyzing affective patterns that can be commonly extracted from people's emoticon usage, another body of research focuses on the contextual and personal aspects, to uncover how people assign new meanings to emoticons: \eg, repurposing or appropriating emojis beyond their originally intended meanings \cite{Wiseman2018_repurposingemoji,kelly2015characterising}, expanding nonverbal expressions to reduce the dependency of texts \cite{zhou2017_wechat}, or developing highly customized usage \cite{Griggio2021}.

Studies encompassing emoticons and emotional communication often used theories and models of emotion as their basis for measures and analyses. Overall, two types of theories were most frequently referred to: discrete emotion theories, which generalize emotions into discrete categories \cite{ekman1992there}, and dimensional emotion theories, which model emotions in continuous dimensions: \eg, valence-arousal model proposed by Russel and Barrett \cite{russell1999core}. For instance, Rodrigues \etal\space established the Lisbon Emoji and Emoticon Dataset (LEED) \cite{rodrigues2018lisbon}, in which they collected 505 participants' perception of 238 emoticons, including the dimensions of valence (positive---negative), and arousal (arousing--calm). These emotional properties of emoticons could inform both the analysis of emoji usage and the development of emotional interfaces. LEED was also used in this study in developing a recommendation algorithm for multi-modal emoticons.

\subsection{Emerging Multi-modal Emoticons in Current Social Applications}

\begin{table*}[tb]
  \caption{An overview (nonexhaustive) of emerging multi-modal or user-customizing features of emoticons in current social applications (customization refers to whether users could modify the properties of emoticons, rather than importing new sets of stickers).}
  \vspace{-3mm}
  \label{tab:socialapplications}
  \small
  \begin{tabular}{cccccc}
    \toprule
    Application &Animation &Haptics &Recommendation &Customization &Customize On-the-fly\\
    \midrule
    Facebook Messenger & $-$ & $-$ & $-$ & $-$ & n/a\\
    Apple Memoji/Animoji & $++$ & $-$ & $+^{\space\space\textbf{1}}$  & $++^{\space\space\textbf{3}}$ & $-$\\
    Apple iMessage Effects & $++$ & $++$ & $-$ & $-$ & n/a\\
    Google Gboard & $+$ & $-$ & $+^{\space\space\textbf{1}}$ & $+^{\space\space\textbf{4}}$ & $+$\\
    Telegram Messenger & $++$ & $+$ & $-$ & $-$ & n/a\\
    Tencent WeChat & $++$ & $+^{\space\space\textbf{2}}$ & $+^{\space\space\textbf{1}}$ & $-$ & n/a\\
    Huawei MeeTime & $++$ & $+^{\space\space\textbf{2}}$ & $-$ & $-$ & n/a\\
  \bottomrule
  \multicolumn{6}{l}{\textbf{Notes}:} \\
  \multicolumn{2}{l}{$++$ : fully implemented;} & \multicolumn{2}{l}{$+$ : partially implemented;} & \multicolumn{2}{c}{$-$ : not implemented}\\
  \multicolumn{3}{l}{$^{\space\textbf{1}}$  Emoji recommendation based on current text input.} &
  \multicolumn{3}{c}{$^{\space\textbf{2}}$  repetitive and simple vibration.} \\ \multicolumn{3}{l}{$^{\space\textbf{3}}$  pre-configured avatars and pre-recorded animations.} &
  \multicolumn{3}{c}{$^{\space\textbf{4}}$  feature mash-up of two Emojis.} \\
  \end{tabular}
\end{table*}

While emoticons have gained enormous popularity in social communications, the traditional, static emoticons seem no longer to suffice people's various needs for self-expression. Users continue seeking more customized, further enriched forms of emoticons. 
Many social applications have developed their unique sets of emoticons (\eg, Facebook Messenger \cite{facebookmessenger}, Telegram Messenger \cite{Telegram}), and a few of them have also explored certain user-customization features (\eg, Apple Memoji \cite{applememoji}, Google Gboard \cite{googlegboard}). More importantly, current social applications start to support (yet limited) multi-modal features in emoticons (see \autoref{tab:socialapplications}).
These emerging features of popular social applications suggest that emoticons are evolving from simply static visual pictograms into more dynamic compounds of multi-modal elements. 
Meanwhile, we could expect such multi-modal emoticons to afford more customization space for users to create richer and more engaging ways of expressing themselves and empathizing with others. 

However, as \autoref{tab:socialapplications} summarizes, current social applications have not fully leveraged the potentials of user-customization in multi-modal emoticons, leaving several meaningful opportunities for new design explorations. 
First, while animation and haptic feedback are increasingly incorporated, current systems only enable pre-designed, fixed combinations of animation and vibration, either applied on specific emoticons (\eg, WeChat \cite{tencentwechat}, MeeTime \cite{huaweimeetime}), or provided as rigid options (\eg, iMessage Effects). None of the systems allow users to freely combine different animations, vibrotactile patterns, and emoticons to create new meanings. 
Second, current systems (\eg, Memoji \cite{applememoji}, Gboard \cite{googlegboard}, WeChat) only explored the recommendation of emojis based on users' text input. None of the existing cases has explored how to recommend multi-modal combinations for users. 
Third, few systems have been designed to facilitate users' customization on the fly. Memoji and Animoji, for example, both require users to pre-define the avatar or pre-record animations to prepare customized stickers, instead of spontaneously creating and sending new multi-modal emoticons according to the unfolding conversation. Although Gboard supports users' mash-up of two Emojis during a conversation, it does not enable on-the-fly modification of animations or haptic feedback. 
 
The above opportunities have underlain the core design features of the \name{} system, which (1) supports users to freely combine various emoticons, vibrotactile patterns, and animations as multi-modal emoticons, (2) provides appropriate recommendations about relevant multi-modal elements, and (3) enables users' on-the-fly customization during a conversation. By embodying these design features that extend beyond existing cases, and evaluating the design in the wild, we aim to extract new empirical knowledge that could advance the development and user experience of multi-modal emoticon systems.

\subsection{Novel User Interfaces Supporting Communications via Emoticons or Multi-modal Signals}

\label{literature_study}

To the best of our knowledge, prior work has rarely explored supporting users to author and communicate via multi-modal emoticons. However, rich novel interfaces have been created and studied in the HCI community, to facilitate users' communications via (visual) emoticons or multi-modal signals (\eg, haptic experience). These design cases have served as inspirations for our exploration.

A stream of research has been focused on exploring new ways for users to communicate via (pictorial) emoticons or emojis. Opico \cite{Khandekar2019_emojifirst} enables emoji-first communication: users could respond to each other using sequences of emojis, expressing feelings or simple concepts without texts. MojiBoard \cite{alvina2019mojiboard} is a keyboard that eases users' entry of parametric emojis: a series emojis that can convey emphasis or micro-stories. 
A number of cases targeted fully automating users' process of selecting or sending emojis. ReactionBot \cite{Liu2018_reactionbot} captures users' facial expressions and accordingly adds emojis to their text messages on Slack. 
Another face-to-emoji idea was explored in \cite{Ali2017_face2emoji}. Other explorations concerned automating the selection of emojis/emoticons based on emotion keywords \cite{Urabe2013_emoticonreco}, sentences \cite{Kim2015_pictogramgenerator}, or speech signals \cite{Hu2019_speechtoemoji}. Voicemoji \cite{Zhang2021_Voicemoji} explored voice-based emoji entry for visually impaired users. 
A few other studies also tackled challenges related to the accessibility and inclusiveness of emoticons \cite{Tigwell2020_emojiaccessibility,Kimura-ThollanderKumar2019_Culturalemoj}. Increased attention has been paid to user customization of emoticons \cite{Griggio2019}, for example, in generating new emojis based on users' sketch and text input \cite{Mittal2020_photorealisticemoji}, or allowing two intimate users to co-customize their emoticon shortcuts (DearBoard) \cite{Griggio2021}.

A few studies suggested the promises of integrating animated representations to enlarge the communication capacity of (static) emoticons. For instance, Animated GIFs were found to afford rich interpretation and nuances in nonverbal communication \cite{jiang2017_GIF}. However, in GIFs/short videos, the image and motion effects are fixed combinations, whereas in this study, we aim to enable users to freely combine static emoticons with different motion effects. As related, the AniSAM study suggested that adding animated representation to static icons could more effectively visualize emotional states (\eg, arousal) \cite{Sonderegger2016_AniSAMAniAvatar}. this suggests adding animated effects might enrich the nuances and expressiveness of static emoticons. Harrison et al. \cite{Harrison2011} proposed Kineticons, a rich library of kinetic behaviors that can be applied on static GUI elements such as icons, to enable extra communication affordance. Our study utilizes Kinneticons to establish a versatile set of animation components for multi-modal emoticon authoring.

Touch, or haptic experience, is an essential type of non-verbal cue for social communications. Besides the abundant haptic studies on supporting users to understand, monitor and operate technologies (\eg, \cite{enriquez2003hapticon,Cauchard2016,Yatani2009}), prior work has extensively explored technology-mediated haptic experience in interpersonal communications. For instance, vibrotactile feedback was leveraged in both mobile and wearable devices, for affective communication in distance, by resembling the feeling of touch, cheek-poke, or handshaking between users: \eg, \cite{Brave1997,VisualTouch,Poketouch,Nakanishietal2014_handshaking,Wang2012_keepintouch}. A recent case explored supporting mediated touch in face-to-face settings without breaking social distancing rules during COVID \cite{Zhang2021_SansTouch}. Several studies have also encompassed supporting haptic communication in text messaging \cite{Mullenbachetal_friction,Israr2015}. In addition to end-user interfaces, related research has contributed to theories \cite{KimSchneider2020}, tools \cite{Chang2020_kirigami,Seifi2015}, or evaluation techniques \cite{Schneider2016} of haptic experience to better support related design and research. For example, VibViz \cite{Seifi2015} offers a large, diverse library of vibrotactile stimuli for haptic feedback design, based on which we developed the vibrotactile library for \name{}. 
In summary, the prior studies illustrated the values of haptic modality in enabling affective communication and social connectedness \cite{Mullenbachetal_friction,Wang2012_keepintouch,Seifi2018}; and a few specifically suggested benefits of combining haptic signals with visual cues \cite{VisualTouch}. These conclusions supported our decision of leveraging haptic feedback as one of the major components for multi-modal emoticons.
\section{\name{} Interface Design}

In this section, we present the design considerations, a usage scenario, and detailed features of the \name{} interface.

\subsection{Design Considerations}
The design rationales behind \name{} are underlain by our explorative research objectives: to extract knowledge about how people create, share and experience multi-modal emoticons in daily communication, and how to better design user-authoring multi-modal emoticon systems. Therefore, as addressed in \autoref{sec:background}, the core features of \name{} are designed to extend beyond, or differentiate from existing design cases, to better facilitate the generation of new understandings in this domain. Here we briefly consolidate our design considerations for these core features, before going into the details of the interface design: 

\textbf{D1: Enabling users to freely combine multi-modal elements.} 
Existing systems that support multi-modal emoticons only provide a limited number of fixed vibration-animation effects for users to customize their emoticons. To probe how to support deeper customization and richer expressions by users, we design \name{} as an open-ended authoring interface that allows for freely combining various emoticons, vibrotactile patterns, and animation effects. 

\textbf{D2: Providing recommendation without limiting users' options.} 
Current emoticon systems have explored automating users' selection by recommending an emoticon-based on their textual inputs (keywords). However, to enable users' nuanced, differing expressions, \name{} explores an alternative recommendation approach that facilitates, rather than fully automated, users' choice. Instead of recommending the best solution, the interface prioritizes the most relevant multi-modal options based on the users' present selected element, while still keeping all other options available. This supports both the user's efficient authoring of classic combinations and open exploration of new expressions. 

\textbf{D3: Supporting on-the-fly authoring of multi-modal emoticons.} 
The user-customization process of existing emoticon systems is often carried out separately from the actual use during online chatting. Users often need to pre-configure emoticon features or dynamic effects in order to use them later. \name{} is designed to support users' on-the-spot authoring of multi-modal emoticons during online communication. It streamlines users' authoring process into four steps: selecting three multi-modal elements and press the ``send'' button (see \autoref{fig:teaser} (b)). With \name{}, we are able to probe how users may create ad-hoc, improvised expressions based on the unfolding conversation.

\subsection{Usage Scenario and \name{} Interface} \label{sec:scenario}

Here we present an example of how users could use the \name{} mobile application to combine emoticons, vibrotactile patterns, and animation effects to create multi-modal emoticons on the fly, during communication. 
Anna and Brad are chatting online. Brad said something really funny which has made Anna burst into laughter. Anna wants to fully express how much she enjoyed it, so she decides to send a ``laughing tears'' emoji \wsicon{images/emoji_laughing_tears} and enhance its expression with animation and haptic effects. She unfolds the \name{} keyboard, and selects that emoji, as well as a vibrotactile pattern that feels like body shaking, and animation of bouncing up and down. Anna is satisfied with this combination, so she presses the ``send'' button, and this emoticon enhanced by the animation effect and vibrotactile pattern is sent to Brad's device. Brad receives it and knows that Anna finds what he said funny. He wants to let Anna know that he feels the same and is still laughing about it. So he pressed that emoticon from Anna, and its haptic feedback is rendered again with the animation on both of their devices. Anna feels this response from Brad. 

As \autoref{fig:teaser} (a) shows, the \name{} interface looks similar to an Emoji keyboard, and can be easily unfolded during messaging by pressing a small icon next to the text-input field. But different from a conventional Emoji keyboard which only shows emoticons, the \name{} keyboard has three segments to display three types of multi-modal elements: emoticons, vibrotactile patterns, and animation effects (\textbf{D1}). As \autoref{fig:teaser} (b) shows, each segment could be scrolled horizontally to browse its all elements. 
Users could make selections from the three sets of elements regardless of the selection order (\textbf{D3}), \eg, they could start from an emoticon like Anna did in the scenario, but they could also start with selecting a vibrotactile pattern or animation effect first. Users could press an element to select it, and it will be highlighted on the interface (\autoref{fig:teaser} (b)). When pressed again, that element will be deselected. 

In the animation segment, all animation effects loop continually so that users could easily preview them. When no emoticon element is selected, the animation previews will be applied on a default neutral icon; otherwise the animation previews will be rendered using the currently selected emoticon. 
In the haptic pattern segment, each vibrotactile element is displayed by a thumbnail preview that visualizes its waveform (based on its intensity and sharpness parameters) in a simplistic manner. These thumbnail previews are designed to help users quickly see the general trend and length of a vibrotactile pattern. Each time when selected, a vibrotactile pattern will be rendered by the haptic module of the mobile phone. If there is a selected animation, the vibration will be rendered in synchrony with the animation. This helps users directly preview what their current combination feels like and eases their quick experimentation with different options.

To further assist users' seamless creation on the spot, \name{} uses an unobtrusive recommendation technique to predict and prioritize the options that are more relevant to the user's current selection (\textbf{D2}). 
Namely, each time when the user makes the first selection, be it an emoticon, vibrotactile pattern, or animation effect, the display order of the elements from the other two segments will be updated accordingly. The elements that are predicted as more likely to be combined with the current selection, will be put forward on the display, so that they can be more easily found. 

This recommendation follows two-fold principles. First, elements that were frequently combined with the current selection and sent by the user in the past, will be prioritized by the display. On top of that, elements that have more similar perceived emotional properties to the selection (based on their scores in the valence and arousal dimensions \cite{russell1999core}), will have higher priority on the display. 
For example, assume that the user's first selection is an emoticon scored relatively high in both valence and arousal, \ie, perceived to convey both ``positive'' and ``exciting'' emotions. Then, the elements in the vibrotactile and animation segments which convey similarly positive and exciting feelings will be given higher priority on the display, to better match the present communication intention of the user. 
This way, \name{} offers a non-aggressive recommendation approach: on the one hand, it eases users' selection by prioritizing options that are more often used or have more emotional relevance to the preceding selection; on the other hand, it still guarantees users abundant space to explore and experiment with new multi-modal combinations. The input data and algorithm of the recommendation technique will be detailed in the next section.
\section{\name{} System}

In this section, we report how \name{} is developed to support the aforementioned design features, including the preparation of the multi-modal elements and data collection of their emotional qualities, as well as the scheme and implementation of the system. 

\subsection{Preparing Multi-modal Elements: Stickers, Animations, and Vibrotactile Patterns}
\label{sec:elements}

Each \name{} emoticon can be authored by a user by combining three multi-modal elements: a (static) sticker, an animation effect, and a vibrotactile pattern (\textbf{D1}).
These stickers, animation, and vibration sets are intended to be open-ended, meaning that users could add new elements to expand each set (\eg, a user could include new stickers to the sticker set). We have prepared rich default elements for the three sets (50 stickers, 15 animations, and 60 vibrations), which affords a great number of combinations possible for creating multi-modal emoticons (also see diverse examples gathered from the field deployment discussed in \autoref{openended}). 

\textbf{Stickers:}
The default set consists of 50 Apple emojis (Version IOS 10.0). We chose Apple emojis as our default stickers due to their frequent usage among mobile users.
The chosen 50 stickers (see \autoref{fig:teaser}) were selected based on an emoji usage survey by the Unicode Consortium\footnote{https://home.unicode.org/emoji/emoji-frequency/}, which classified all the Unicode emojis based on usage frequency. We chose the frequently used facial emoji stickers (\eg, not objects, stars, etc.). This set of emoticons has also been targeted in Rodrigues \etal's survey \cite{rodrigues2018lisbon}, due to their usefulness and frequent usage for expressing a variety of emotions. 

\textbf{Animations:}
The default set consists of 15 animation effects from the work of Kinecticons \cite{Harrison2011}. It provides a set of diverse, open-ended and multi-purpose kinetic effects to be combined with different graphical elements (\eg, icons) to support a wide range of communication purposes \cite{Harrison2011}. Hence, these animations are not strongly bonded to certain semantic meanings or connotations. This open-endedness is in accordance with our design, which is to combine the animations with different stickers to afford rich expressions. 
Of the original 39 Kinecticons, we excluded the animation effects concerning two objects, which are not applicable in our case. We also filtered out those that could not be appropriately applied on circular emoji stickers, \eg, because they were designed for  rectangular icons or menus.

\textbf{Vibrations:} The 60 default vibrotactile patterns were chosen from the work of VibViz \cite{Seifi2015}, a diverse vibration library intended for the design to convey information through the haptic channel in digital devices. VibViz was chosen also because of its potential possibilities to be combined with different stickers and animations to convey various meanings.
Of the original 120 vibrations, we excluded the ones that were longer than 10 seconds, because they are not suitable for short, emoticon-based instant-messaging scenarios.
We further filtered out vibrations that had less emotional relevance to the default set of stickers in terms of valence and arousal (further explained in \autoref{vibration_data_collection}).

\subsection{Gathering Data for Perceived Emotional Properties of Animation and Vibrotactile Libraries}
\label{sec:data-gathering}

As mentioned, our design rationale is to grant great freedom for user authoring while keeping the authoring process intuitive and effortless (\textbf{D1} and \textbf{D3}). To achieve this, we use a recommendation algorithm that predicts most likely combinations based on the selected element, and accordingly updates the display order of the un-selected sets, to ease the choices of users. 
Such prediction is based on two kinds of data: the perceived emotional properties of each element (\ie, valence and arousal \cite{russell1999core}), and the user's historical usage data. 
As the prerequisite of the recommendation algorithm, the datasets of perceived emotional properties need to be established beforehand. For the sticker set, the perceived emotional properties of the core collection of Apple emojis have been already measured by Rodrigues \etal \cite{rodrigues2018lisbon} (with a sample size of 505), and this measured collection has covered the frequently used emojis we chose based on the Unicode survey. We therefore utilized their open-sourced data for the algorithm. However, for the animations and vibrations, adequate datasets remain to be established. We thereby conducted questionnaire surveys to gather the data.

\subsubsection{Data Collection for Animations}
A web-based questionnaire survey was conducted with 52 respondents to gather the perceived emotional properties of the 15 Kineticon animations. Kineticons were designed with a rich vocabulary of kinetic behaviors to convey diverse intentions and meanings \cite{Harrison2011}. However, no prior work has assessed what emotional feelings these animations could trigger from users, which is crucial for emotion-based recommendation.

The questionnaire was specifically developed for data collection using \texttt{Javascript} and \texttt{Node.js}, which renders each animation effect on a separate page and asks respondents to use 7-point Likert scales to rate the valence and arousal of each animation (Supplementary Material A). The animations are rendered using the neutral default icon used in the original user study of Kineticons \cite{Harrison2011}.  
The order of the animations was randomized for each respondent. 

\begin{figure*}[tb]
  \centering
  \includegraphics[width=\linewidth]{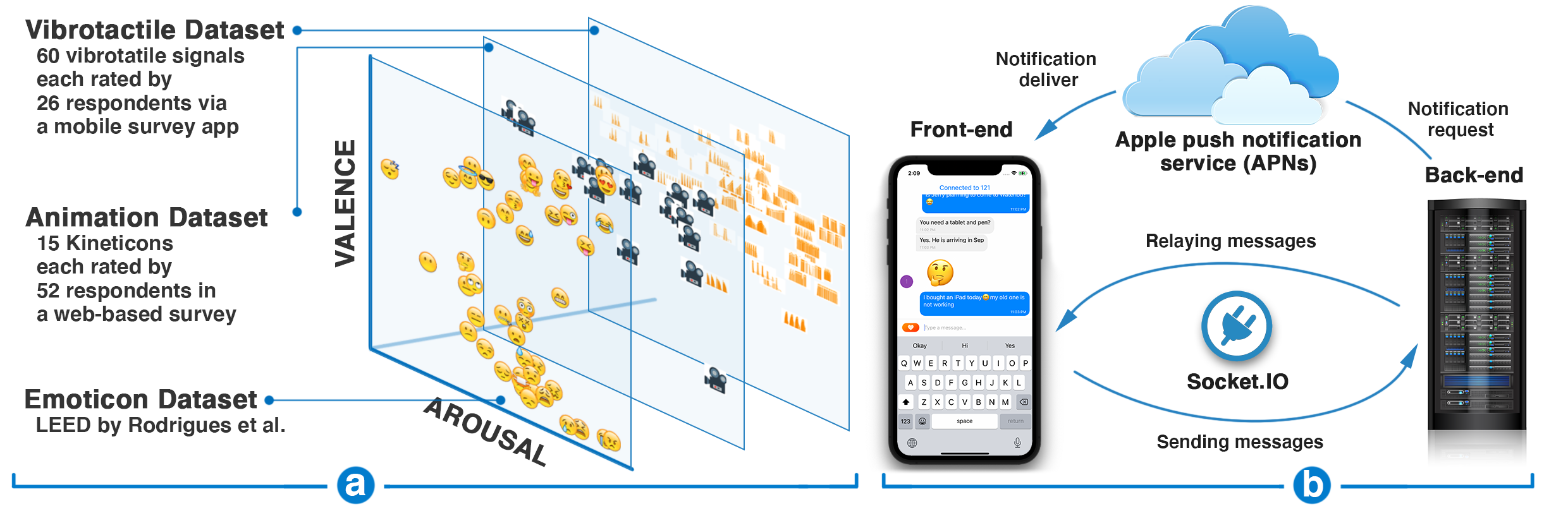}

  \vspace{-4mm}
  \caption{(a) emotional property datasets of stickers, animations and vibrations for providing recommendation based on the elements' emotional relevance; (b) a simplified illustration of the \name{} system architecture.}
  \Description{alt-texts TBA}
  \label{fig:system}
\end{figure*}

The respondents were recruited from the Amazon Mechanical Turk (MTurk) platform (aged from 25 to 54; 27 males, 24 females, and 1 other). Each survey session took 10--20 minutes from a respondent. The compensation for each respondent is \$2.5. 1 invalid responses were excluded due to incompleteness. In the end, 52 responses were used (each of the 15 animations was rated by 52 respondents) as the prerequisite data of the animation set (see \autoref{fig:system}). 

\subsubsection{Data Collection for Vibrations}
\label{vibration_data_collection}
The original VibViz study results consist of the valence and arousal scores of each vibrotactile pattern rated by three researchers \cite{Seifi2015}.
However, no survey has been conducted so far, to collect more data about users' perceived emotional properties of the VibViz patterns, which had become part of our development tasks. 
Besides excluding long vibrations (>10s), we filtered out vibrations that were relatively farther to an emoticon sticker from the sticker set in the valence-arousal space. These more distant vibrations had less emotional relevance to the chosen emoticon stickers, therefore, they are deemed less likely to be combined together with the chosen stickers. 

For each sticker in our sticker set, we marked the 5 closest vibrations in the valence-arousal space, and accumulated the mark frequency for each vibration. In the end, there were 60 variations marked more than once, which resulted in our final vibration element set.

A second questionnaire survey was conducted with 52 respondents for the emotional perception of the vibrotactile patterns.
Due to the fact that respondents need to access a device with a haptic engine to experience the vibrotactile patterns, we developed an ad-hoc survey application on the iOS platform using Swift UI (Supplementary Material B). 
To avoid each survey session getting too long and causing fatigue or boredom in respondents, we split the 60 vibrotactile patterns into two surveys (30 for each). Each survey took 15--20 minutes for each respondent. 
In the survey application, to rate each vibrotactile pattern, the respondent was asked to first press a button to experience the vibration before its Likert scales appeared. This button was available on-screen throughout the rating of that pattern, so that the respondent could replay the vibration anytime when needed.

Due to the consideration of device access, respondents were only recruited through word of mouth, which were required to have access to an iPhone 8 or above (to make sure the haptic engine works smoothly). The survey app was distributed to the respondents through both local build and TestFlight\footnote{https://developer.apple.com/testflight/}. In the end, 52 qualified respondents (aged from 18 to 59; 35 males and 17 females) completed the survey and each vibrotactile pattern was rated by 26 respondents (see \autoref{fig:system}).

\subsubsection{Post-survey Questions}

After rating the emotional properties collected for the recommendation algorithm (\autoref{fig:system}), the 104 respondents of the animation and vibration surveys filled in two extra multiple-choice questions: if you could send emojis combined with animations and vibrations in online communication, (1) what would be the added values that you foresee, and (2) what would be the meaningful usage scenarios for it? The choices of these questions were formulated based on our literature study (\autoref{literature_study}) and pilot interviews with five college students. Open questions were not used to avoid the extra questions demanding too much time and effort from the respondents. 

\begin{figure*}[tb]
    \centering
    \includegraphics[width=\textwidth]{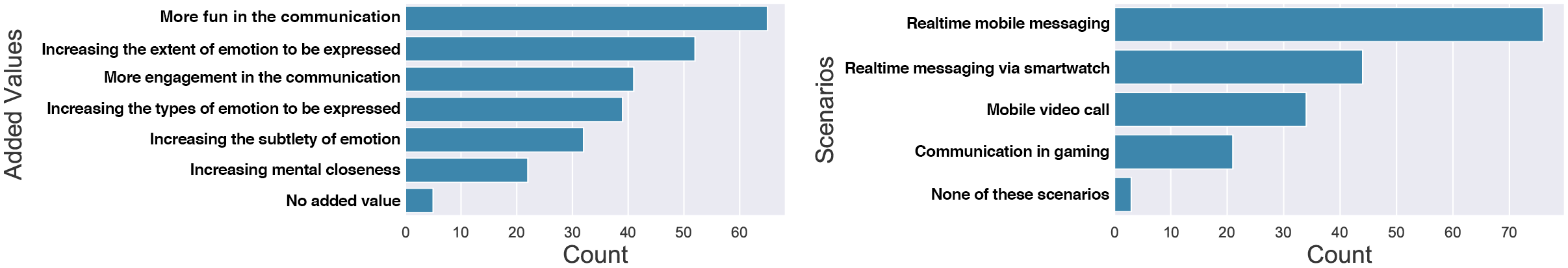}
    \vspace{-6mm}
    \caption{Frequencies of responses on the added values and usage scenarios of sending emojis combined with animations and vibrations.}
    \label{fig:animation-vibration-survey}
\end{figure*}

The results are shown in \autoref{fig:animation-vibration-survey}.
For added values, the most frequent response is more fun in the communication (65 times), followed by a few other answers. 
These responses inspired the questionnaire developed for our field study (see \autoref{sec:fieldstudy-methods}), to see whether similar opinions would emerge from lived experiences with multi-modal emoticons. In terms of meaningful scenarios, the most frequent option is real-time messaging via mobile phone (76 times).
This confirms that mobile messaging is a meaningful scenario to start exploring multi-modal emoticons.

\subsection{Recommending \name{} Elements}

As demonstrated in \autoref{sec:scenario}, to ease the authoring process of multi-modal emoticons, \name{} offers an unobtrusive recommendation to predict and prioritize the multi-modal emoticon elements (\ie, stickers, animations, and vibrotactile patterns) that are more relevant to the user's current selection (\textbf{D2}).
Initially, before any user selection, all the three elements have default display orders on the \name{} interface. This mimics the convention of most emoji keyboards where the location of each emoji is fixed to facilitate users with consistent access based on their memory. 
Once a selection is made by the user, the undecided elements are ranked based on their association to the selected element by the user. Given an undecided element $u$ in one modality $M_u$, and a selected element $s$ in another modality $M_s$, the recommendation algorithm considers the following two essential factors to obtain a ranking score for $u$. 

First, elements that have more similar perceived emotional properties to the selected element are prioritized. This is because it is less appropriate to combine elements that are quite different in their emotional properties, which may result in expressing contradictory emotions in a single \name{} emoticon.
Based on the gathered data described in \autoref{sec:data-gathering}, we simply use the Euclidean distances between elements in the valance-arousal 2D plane, like those in previous studies \cite{Seifi2015,rodrigues2018lisbon}. 
Specifically, the perceived emotional similarity $P$ is defined as:
\begin{equation}
    P(u, s) = \frac{1}{\sqrt{(u_v-s_v)^2+(u_a-s_a)^2}},
\end{equation}
where $u_{v,a}$ and $s_{v,a}$ represent the valence and arousal values of $u$ and $s$.

Second, elements that were frequently combined with the current selection and used in the past are prioritized. This is based on the assumption of recently visited items will be likely revisited again, which is widely used in a range of recommendation systems such as Google Search, Amazon, etc.  
However, simply counting the frequency of $u$ and $s$ used in combinations overlooks the overall usage patterns.    
We thus employ a TF-IDF approach \cite{rajaraman_ullman_2011} that is commonly used in information retrieval. The TF (term frequency) reflects the frequency of $u$ and $s$ used together with respect to $u$ being used with all the elements in the modality $M_s$. That is,
\begin{equation}
TF(u, s) = 
    \begin{cases}
    \frac{F(u, s)}{\sum_{s' \in M_s} F(u, s') } \\
    0, \textrm{if } {\sum_{s' \in M_s} F(u, s')} = 0
    \end{cases},
\end{equation}
where $F(u,s)$ is the number of occurrences that $u$ and $s$ are combined in \name{} emoticons by the user in the past.
Further, the IDF (inverse document frequency) reduces the bias that the element $u$ is used too often with all the elements in the other modality $M_s$, which promotes the diversity of the recommendation. Thus, 
\begin{equation}
    IDF(u,s) = \log \frac{n}{\sum_{s' \in M_s} N(u,s')} + 1, 
    N(u,s) = 
    \begin{cases} 
        0, \textrm{if } F(u,s) < 1\\
        1, \textrm{if } F(u,s) \ge 1
    \end{cases},
\end{equation}
where $n$ is the number of elements in $M_s$.
Therefore, the final TF-IDF score is $TF\text{-}IDF(u,s)=TF(u, s) \cdot IDF(u,s)$.

We then use a weighted score to compute the ranking score for $u$ with respect to $s$ by combining the two factors:
\begin{equation}
    R(u,s) = \alpha P(u, s) + \beta TF\text{-}IDF(u,s),
\end{equation}
where $\alpha$ and $\beta$ are weights. In our development, we set $\alpha=0.6$ and $\beta=0.4$ based on our empirical observation. 
We compute this ranking score for every element $u \in M_u$ and sort the scores descendingly to obtain the display order of the elements on the \name{} interface.
When there are only one selected element (\eg, a sticker), we reorder the elements in the other two modalities (\eg, vibration and animation) based on the above ranking method.
When there are two selected elements (\eg, a sticker and a vibration), we simply use the average of the two ranking scores to obtain the final score of each element in the third modality (\eg, animation). 

\subsection{Implementation Details}
In this section, we introduce how we built the \name{} system in detail. As shown in \autoref{fig:system}, the system contains a front-end mobile application to support on-the-fly authoring of multi-modal emoticons, and a back-end server to perform data processing, messaging, and communication.

The front-end application was implemented based on the React Native framework, because it can be easily adapted to other platforms in the future (\eg, Android), in addition the Apple iOS platform this project has been built upon. 
To have fine-grained control over how the vibrotactile patterns are played on the device, we developed a custom React Native bridge to expose the native iOS haptic API. Our implementation enabled the app to play and pause haptic patterns defined in AHAP (Apple Haptic and Audio Pattern) files, which were identical to the ones used in surveys. 
For the animations, we recreated the selected Kinections (see \autoref{sec:elements}) using \texttt{react-native-reanimated}. 
We used \texttt{react-native-gifted-chat} to implement the front-end messaging application, which provides a comparable look and feel to other messaging applications (see \autoref{fig:teaser}).

The back-end server was developed using \texttt{Node.js}, which is responsible for relaying messages between users and delivering notifications. To increase the robustness of the system, the server stores a queue of undelivered messages when the receiving party is offline, and re-sends these messages once the users are re-connected. Specifically, notifications on iOS are managed using APNs (Apple Push Notification service).
We used \texttt{socket.IO} for sending and receiving messages between the front-end and the back-end for reliable real-time communication. To enable sending and receiving \name{} emoticons, we created a custom encoding in plain text behind the rendered multi-modal emoticons. 
For each emoticon, the front-end sends the encoded string to the back-end which relays the message to the receiving party. Then, the front-end on the receiving side re-renders the message based on the encoded string.

\section{Field study}

Using \name{} as both a design to study about, and a technology probe to study with, we conducted a four-week field evaluation to gather empirical knowledge about how people create, share and experience multi-modal emoticons in daily online communications, and thereby surface relevant design implications for future research and development.

\subsection{Participants and Study Setup}

Twenty participants (in 10 pairs) were recruited for a four-week field evaluation. The participants were aged 19--36 (\md{29}, \iqr{10.5}), with 12 females and 8 males. They are residents from three countries: Canada (10 participants), China (8 participants), and the Netherlands (2 participants). 
Each pair of participants had already established a social relationship with each other; namely, they had known each other for 10 months to 22 years (\md{4 years}, \iqr{2}). 
Participants are referred to as P\# in the following text, and consecutive numbers indicate the pairs (\eg, P1 \& P2; P3 \& P4).
Among the 10 pairs, nine of them were friends, and one pair was significant others (P15 and P16). More details about the participants could be found in Supplementary Material C. Participants were recruited via the word of mouth based on two basic requirements: (a) they already use an iPhone 8 or above as the primary mobile phone so that the built-in haptic engine ensures the same sensorial quality of the designed vibrotactile elements; and (b) they can be paired with another person that they already know and have regular communications via online applications.

In the beginning of the study, each participant installed the \name{} mobile application on their mobile phone through Apple TestFlight. To ensure that they could experience the system in a naturalistic manner, they were asked to use the application in their daily lives to communicate with each other as they already did prior to the study. No structured sessions of chatting were required by the researchers. Neither were the participants asked to use the system exclusively: they were encouraged to use \name{}, while they could still use their existing online communication tools at the same time. It was reminded that some risky information might need to be avoided when using the application to chat, such as bank accounts or passwords. Some common and safe topics were provided as examples: including updates about each other's recent life, recent news from the public media, or planning future activities together. Meanwhile, it was made clear to the participants that these were just examples, and they could discuss any topics they desired. Each participant was compensated \$15 after their participation.

\subsection{Data Gathering Methods} \label{sec:fieldstudy-methods}

Given our highly explorative goal, we adopted a mix-method approach to gather both quantitative and qualitative data in various forms during the field evaluation period. The details are as follows.

\textbf{Unstructured feedback.}
During their whole period of participation, the participants were encouraged to share their experience and thoughts to the researchers anytime as desired, through multiple types of communication channels, such as voice calls, voice messages, or textual messages. Along with such feedback, they also used screenshots or screen recordings to share some usage examples of multi-modal emoticons that they found meaningful or interesting. These quick and unstructured feedback sessions were transcribed and annotated for analysis. 

\textbf{System logs.}
Participants' usage logs stored in the back-end were partially used for generally understanding how they interacted with the interface. These retrieved system logs included the number of messages or \name{}'s they had sent, as well as their operations on the \name{} interface: \eg, when and what they pressed on the interface while selecting the multi-modal elements to construct a \name{}. To be noticed, the textual contents of participants' chat history were not directly retrieved by researchers for analysis, unless the participants proactively shared a certain part of their chats in unstructured feedback or in final interviews. 

\textbf{Questionnaire.}
Towards the end of their participation, each participant was provided with an online questionnaire via Google Forms, asking about their general experiences of using \name{}. In the questionnaire, we asked about their general styles of constructing multi-modal emoticons: \eg, in which order they made selections across the three types of multi-modal elements, and whether they had discovered several classic combinations and reused them in chat. 
Moreover, we asked about their general experience of \name{} in comparison with the conventional emoticons, in terms of conversation engagement, fun, expressiveness, etc. (see \autoref{fig:quan_results}). 
These data were meant to gather their high-level experience and general opinions which could serve as supplementary or triangulation to the in-depth interview data.

\textbf{Semi-structured interview.}
A semi-structured interview was conducted for each pair, to gather vivid, in-depth empirical data about how they used \name{} to create and share multi-modal emoticons, and what their detailed experiences were. Each interview took 30--45 minutes and was structured in three sections. 
\begin{itemize}
    \item \textit{Section 1 (8 minutes):}
    The pair was asked to describe their general experience of using \name{}. They were also asked to give more explanation about the general experiences they rated in the final questionnaire.
    
    \item \textit{Section 2 (15 minutes):}
    The pair was asked to provide concrete examples about the multi-modal emoticon they created and considered meaningful in the specific context. In addition, they were asked to reflect on their detailed workflows, such as how they selected different elements to construct new multi-modal emoticons, and how they experience the various design features.
    
    \item \textit{Section 3 (7 minutes):}
    To probe their latent needs about multi-modal emoticon systems, the pair were asked to envision what important features the next version of \name{} should have. To probe future design opportunities, the pair was asked to envisage what future scenarios can benefit from multi-modal emoticons, in addition to instant messaging. 
\end{itemize}
In the end, an open discussion is encouraged for participants to share additional thoughts.
All interviews were audio-recorded and transcribed verbatim for a thematic analysis along with the unstructured participant feedback data.
\section{Results and Analysis}

\subsection{Quantitative Results}

\subsubsection{Questionnaire Responses}
\begin{figure*}[tb]
  \centering
  \includegraphics[width=\linewidth]{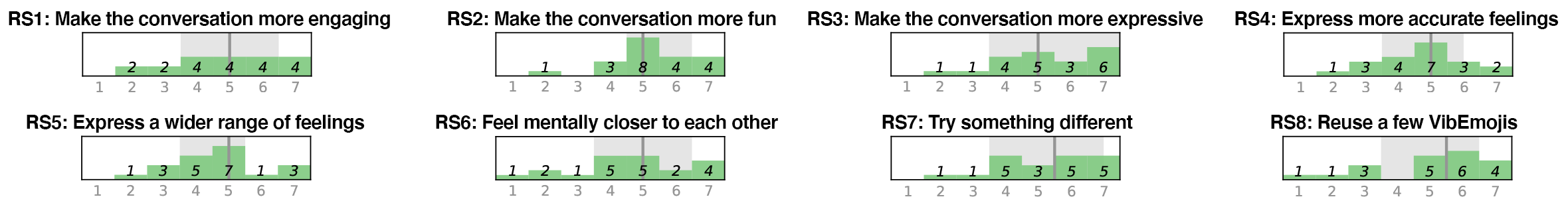}
  \vspace{-7mm}
  \caption{RS1 to RS6: users' general experience with multi-modal emoticons in comparison with static emoticons; RS7 and RS8: users' general style in authoring multi-modal emoticons (7-point Likert scale, 1=strongly disagree, 7=strongly disagree). 
  }
  \Description{alt-texts TBA}
  \label{fig:quan_results}
\end{figure*}

RS1 to RS6 in \autoref{fig:quan_results} show the 20 participants' overall experiences with the \name{} emoticons, in comparison with the static emoticons they had been using before (on a 7-point Likert scale).
In general, they considered that the multi-modal emoticons had contributed to the engagement (RS1), fun (RS2), and expressivity (RS3) of the conversation, and helped them to express more accurate (RS4), and a wider range of feelings (RS5), and increased their mental closeness (RS6) to each other: with all medians residing on 5 (\autoref{fig:quan_results}). 
As shown by the detailed rating distribution in the figure, it seems that users gave higher ratings (i.e., 6 or 7) towards multi-modal emoticons' contributions to the engagement, fun, expressivity, and mental closeness, in comparison with their contributions to the accuracy or the range of feelings.

Moreover, \name{} affords great freedom for users to combine various multi-modal components. To generally probe whether such freedom of exploration was meaningful, the questionnaire asked about their general style of usage: \ie, RS7 and RS8 in \autoref{fig:quan_results}.
As the figure shows, overall, the participants tended to try something different (variations of combinations) when authoring multi-modal emoticons in chatting (RS-7: \md{5.5}). Meanwhile, they also tended to reuse a few combinations (RS-8: \md{5.5}). 
This suggests that while being explorative in use, the participants had also formulated a few frequently used expressions (see examples in \autoref{openended}), which verifies our design choice for supporting both frequency-based recommendation and open exploration (\textbf{D2}).

The open-ended design of \name{} allows users to flexibly decide the order of selecting multi-modal elements. When asked about participants' habituated order, 19 preferred stickers to be the first element to choose, while only one preferred starting from vibration. Divergence emerged among the users, in terms of the second element they preferred to select. While 13 participants would go for animations, six would select a vibration, and the other would go for stickers. More analysis of the users' selection order and decision-making in the authoring process is discussed in \autoref{authoring_process}. 

\subsubsection{Interaction Logs}
During the whole period of the field evaluation, the participants sent 1,824 textual messages and 581 multi-modal emoticons using \name{}: approximately every three text messages were accompanied with one \name{}. 
Per pair of participants, the median number of textual messages sent is 86 (\iqr{162}, \mini{21}, \maxi{954}), and for multi-modal emoticons, the median is 47.5 (\iqr{49}, \mini{15}, \maxi{182}). The logs also provided a rough insight into the time spent for the users to author a multi-modal emoticon. We defined an interaction timeframe that started from the users' first operation on the authoring interface to the moment of pressing the ``send'' button. The median of the interaction timeframes is 7.09 seconds (\iqr{13.01}).
A slight difference could be seen, when comparing the timeframes of the first 10\% (\md{9.6 seconds}, \iqr{18.1}) \name{} emoticons sent with the later 90\% (\md{7.1 seconds}, \iqr{12.7}) across all pairs. This may indicates a learning stage in the beginning, and the users became faster after familiarized with the interface. Moreover, given that users had been actively exploring different variations throughout the study (RS7 in \autoref{fig:quan_results}), these timeframes should also reflect their time spent for exploration and experimentation.

\subsection{Qualitative Results}
A thematic analysis was conducted to analyze the qualitative data gathered from the semi-structured interviews with the participants, as well as their unstructured feedback provided to us throughout the field study. This inductive analysis method was chosen due to our purpose of establishing a set of structured, systematic meanings (themes) \cite{Braun2012} about how the participants created, shared, and experienced multi-modal emoticons in the wild with the \name{} system, in order to generate rich empirical knowledge for future research and design. 
Our analysis followed the six-phase procedure detailed by Braun and Clarke \cite{Braun2012}, which included familiarization with data by organizing notes and annotations across the dataset, generating initial codes based on research objectives, searching for themes by establishing connections among codes, reviewing and finalizing themes, and re-contextualizing the themes to formulate findings. 
In this section, we address these qualitative findings, in light of some quantitative results reported in the prior section. 

\subsubsection{Theme 1: Four Types of Scenarios that Multi-Modal Emoticons Bring Extra Values to Communication} \label{openended}
This theme presents vivid usage examples to illustrate what multi-modal emoticons were created and shared in the wild and how they meaningfully benefited the participants' daily communication experiences. We thereby contextually examine the added values of multi-modal emoticons in addition to conventional (static) emoticons, to offer new empirical understandings about multi-modal emoticons.

In general, the participant agreed that using multi-modal emoticons in their daily communications had contributed to their online chatting experience. For instance, as P16 experienced, \qt{overall, it is pretty nice, it's a good extension of [a current social application]. In real-life, human emotions are multilayered, and our facial expressions are very dynamic [...] These different options [vibrations and animations] could deliver different emotions. It's richer [than conventional emoticons]. It adds more dimensions for communicating emotions.} Similarly, as P5 felt, \qt{It [\name{}] is convenient, easy to use. With vibrations and animations added, the emoji feels more three-dimensional. You could feel more emotions from it.}
It has also been suggested that both heavy users and less-frequent users of existing emoticons, could perceive values of using multi-modal emoticons: \eg, \qt{I send a lot of stickers every day, this [\name{}] could expand my expressions of emotions} (P3); and \qt{I don't use a lot of emojis when chatting, but this type of emojis [multi-modal emoticons] could express richer feelings. Because the emoji has animation and vibration, it has more [information] in it.} (P19). Correspondingly, the participants reported abundant usage cases, and identified different types of meaningful uses of multi-modal emoticons.

\textbf{Scenario Type (i): Enhancing Existing Meanings of Static Emoticons.} The participants provided examples that they used animation effects and vibrotactile patterns to enhance or animate the existing meaning of an emoticon, and made it more vivid or expressive in the conversation. 

For instance, as \autoref{fig:qual_results} shows, P3 once asked P4 if she went to sleep already. P4 said not yet, but she would sleep soon, and then she sent a multi-modal emoticon with the \textit{sleepy face} Emoji \wsicon{images/emoji_sleepy_face}. To enhance this Emoji, she used an animation and a vibration that both could be associated with the behavior of sleep: the animation is spinning slowly and slightly, and the vibration \qt{felt like snoring.} Both P3 and P4 thought that this combination was expressive and made the conversation more fun. The next example is from P15: when chatting with P16, she augmented the Emoji \textit{face blowing a kiss} \wsicon{images/emoji_face_blowing_kiss} with vibrations that felt like the sounds of kissing but had different duration. She further explained that she would use a shorter vibration to represent a light kiss, and a longer vibration for a long kiss. 
As another example, P7 sent P8 a \name{} based on the emoji \textit{smiling face with heart-eyes} \wsicon{images/emoji_smiling_face_with_heart_eyes}, to express the excitement of seeing a celebrity. To enhance the expression, she used the animation of rapid shrinking and expanding, and a vibration felt like heartbeats. 

\textbf{Scenario Type (ii): Creating New Meanings beyond Original Emoticons.} In these examples, the participants flexibly combined the multi-modal elements to convey new meanings that were additional to, or different from the original meaning of the used emoticons. And oftentimes, such new messages or expressions cannot be simply conveyed by the base emoticon alone. 

One example was reported by P12, in which he asked P11 to play an online game but did not get a response. He then sent a multi-modal emoticon combining an animation of rapidly waving from side to side, with the Emoji \textit{grinning squinting face} \wsicon{images/emoji_grinning_squinting_face}, and a vibration that matched the animation. He selected this \qt{vibrant} motion in order to \qt{urge him [P11] to reply}, whereas the Emoji was not selected for specific reasons and just to afford a positive connotation. Another creative case was shared by P16 (\autoref{fig:qual_results}). He appropriated the meaning of the Emoji \textit{zipper-mouth face} \wsicon{images/emoji_zipper_mouth_face} by combining it with a \qt{naughty} and rapid motion, and an intense vibration, which both felt like \textit{}{"struggling}, to express that he did not want to stop talking. As he explained, \qt{This emoji [zipper-mouth face] originally means 'shut up', but [with the selected animation,] now it has become 'I don't want to shut up'. It then had extra meaning.} 
More such examples were offered by the participants. For instance, P17 combined the Emoji \textit{hugging face} \wsicon{images/emoji_hugging_face} with a motion of bouncing up and down, to convey a \qt{creepy} impression when joking with P18. P2 combined the Emoji \textit{grimacing face} \wsicon{images/emoji_grimacing_face} with an animation and a vibration that felt like \qt{trembling with cold} when discussing the weather with P1. 

Above examples demonstrate how users' could creatively construct new expressions and assign new meanings in daily communications when they can freely combine multi-modal emoticons, which confirmed our \textbf{D1}.

\textbf{Scenario Type (iii): Setting Atmospheres for Unfolding Conversation.} Participants found they could construct multi-modal emoticons to (re-)set the atmosphere for a conversation session. In these examples, the prior sent multi-modal emoticons which left on the chat display continued contributing to the ambience of the communication, and reminding conversation partners of certain vibes during a chat session. 

A concrete example was from P15. She had a particular multi-modal emoticon for starting a joke, which included the emoji \textit{winking face with tongue} \wsicon{images/emoji_winking_face_with_tongue}, an animation of zooming in and out, and an intense vibration (\autoref{fig:qual_results}). As she described, \qt{Every time I want to make fun of [P16], or joke with him, I will send this. It's like an opening, so that he won't get mad at me. It sets the tone for the chat [...] a reminder that I'm joking.} 
P9 also mentioned that he used the multi-modal emoticons to create relaxing vibes: \qt{I sent \name{} [emoticons] because I don't want the conversation to be formal or distant. I hope to have some humor, or closeness in the atmosphere.} Such continuous rendering of a certain atmosphere was partially attributed to the animation components of the multi-modal emoticons, which were continuously visible to users as long as the emoticon is displayed. As P7 reflected, \qt{the animation itself created a sort of atmosphere [...] It's playful [...] hard to describe, but it's contagious from person to person. We both [P7 and P8] could get it.} 

\textbf{Scenario Type (iv): Constructing Attentive, Empathetic Responses.} regards constructing multi-modal emoticons to formulate attentive, empathetic responses. In these examples, the participants created multi-modal emoticons as spontaneous responses to their partners in the unfolding conversation. As experienced by their conversation partners, these responses constructed on the spot could convey more attentive and empathetic feelings than responses via simply standard or preset emoticons. 

P8, for example, was once listening to P7 talking about her work. During listening, she also responded to P7 by sending two multi-modal emoticons that both included the emoji \textit{thinking face} \wsicon{images/emoji_thinking_face}, but had different animations and vibrations which all felt relatively slight and calm (\autoref{fig:qual_results}). As she explained, compared with using a static emoticon, or sending the same multi-modal emoticon twice, the spontaneous variations she created on the fly could better convey the meaning \qt{I am listening with care}, since \qt{it shows that I have received your messages and I give you different response each time, rather than simply copying [the same responses].} And this meaning was indeed conveyed to P7: \qt{[P8] sent me the stickers of thinking, with slowly flipping or spinning [...] felt like she was still thinking and thinking about different things I said [...] She was expressing 'I am listening' and she did not want her expression to be interrupting [...] compared with static stickers, this response is more interactive, instead of repeating. It creates richer nuances and feels more friendly.} 

Another example was from P3 and P4 discussing the surgery of P4's father. They both sent a \name{} emoticon built upon the same emoji \wsicon{images/emoji_crying}. What made them feel empathized with each other was that they also used a very similar vibration to express sadness: \qt{It felt nice that we used the same vibration [...] we shared the same feeling [...] we both thought this vibration could express sadness.} Such empathetic experience was also reported by other pairs. For example, P1 and P2 appreciated the \qt{tacit understanding} (P2) or the \qt{the sense of empathy} (P1) between them, when they used similar animation or vibration element to express similar feelings in the conversation: \qt{we could associate with same emotions [...] don't know why but we just can [...] it's not fully conscious.} 
In similar cases, P5 also experienced \qt{resonance} with P6. As put by P8, when responding to each other using multi-modal emoticons with certain similar components, \qt{it feels like we are in sync [...] a sense of being connected.} Above examples suggest the great values of multi-modal emoticons in enriching users' spontaneous non-verbal responses to each other, and enhancing their sense of connectedness and interactiveness in the conversation. 

As shown in the examples, these benefits could only be afforded when users are able to conveniently and flexibly construct multi-modal emoticons on the fly, during the unfolding conversation, which contextually confirmed \textbf{D3}.  

\begin{table*}[tb]
  \caption{Four types of scenarios that multi-modal emoticons suggested extra values to communication experience. Each type is explained by an exemplar case gathered from the participants, with the multi-modal emoticons illustrated. More examples could be found in \autoref{openended}.}
  \centering
  \vspace{-3mm}
  \label{fig:qual_results}
  \small
  \begin{tabular}{m{4.5cm} m{5cm} m{1.5cm} m{5cm}}
    \toprule
    Scenario &Exemplar Case &Vibration &Emoticon and Animation Effect\\
    \midrule
    Scenario Type (i): Enhancing Existing Meanings of Static Emoticons & P4: using a ``snore" vibration to enhance \wsicon{images/emoji_sleepy_face} & \includegraphics[width=1.5cm]{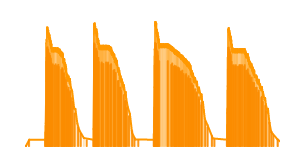} & \includegraphics[width=5cm]{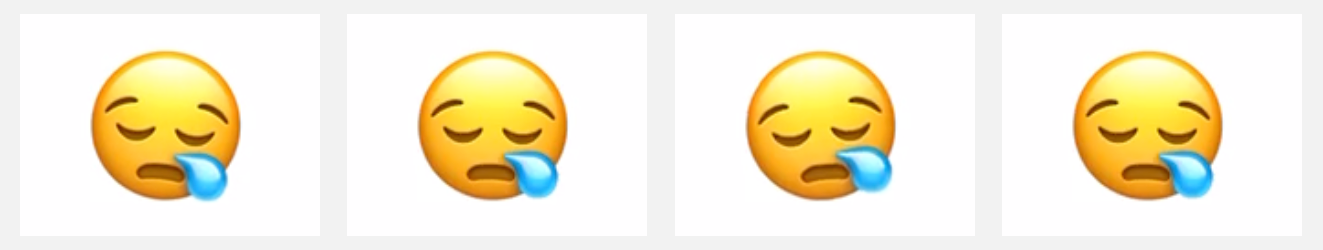}\\
    Scenario Type (ii): Creating New Meanings beyond Original Emoticons & P16: appropriating \wsicon{images/emoji_zipper_mouth_face} with a ``struggling" effect to express ``I don't want to shut up". & \includegraphics[width=1.5cm]{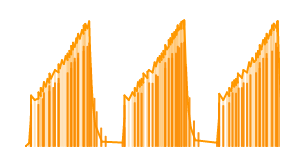} & \includegraphics[width=5cm]{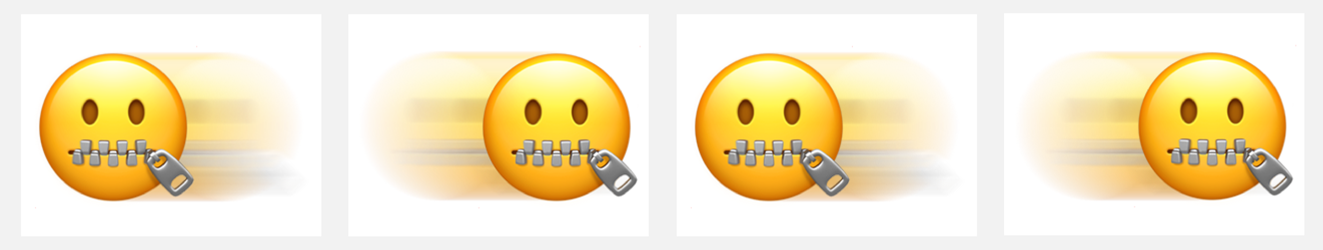}\\
    Scenario Type (iii): Setting Atmospheres for Unfolding Conversation & P15: the multi-modal combination often used to set an atmosphere for making jokes. & \includegraphics[width=1.5cm]{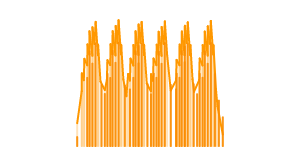} & \includegraphics[width=5cm]{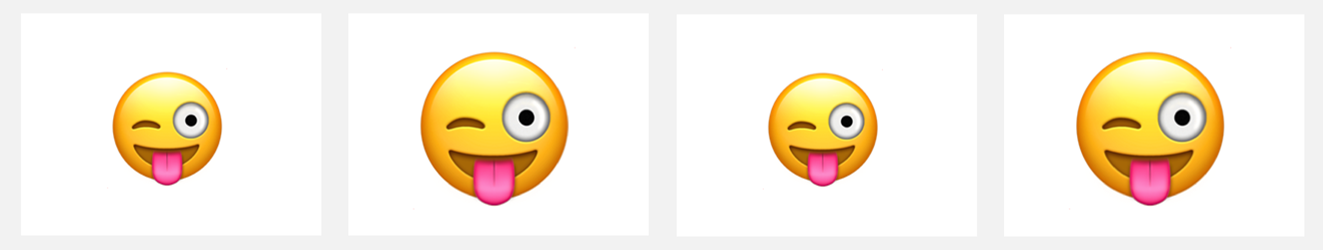}\\
    Scenario Type (iv): Constructing Attentive, Empathetic Responses & P8: the same emoticon with varying motions and vibrations conveyed attentive, non-repeating responses, without interrupting the conversation partner. & \includegraphics[width=1.5cm]{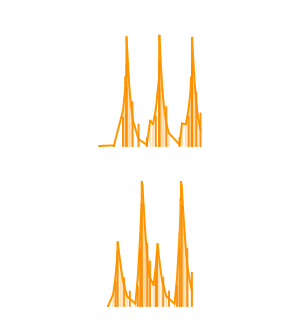} & \includegraphics[width=5cm]{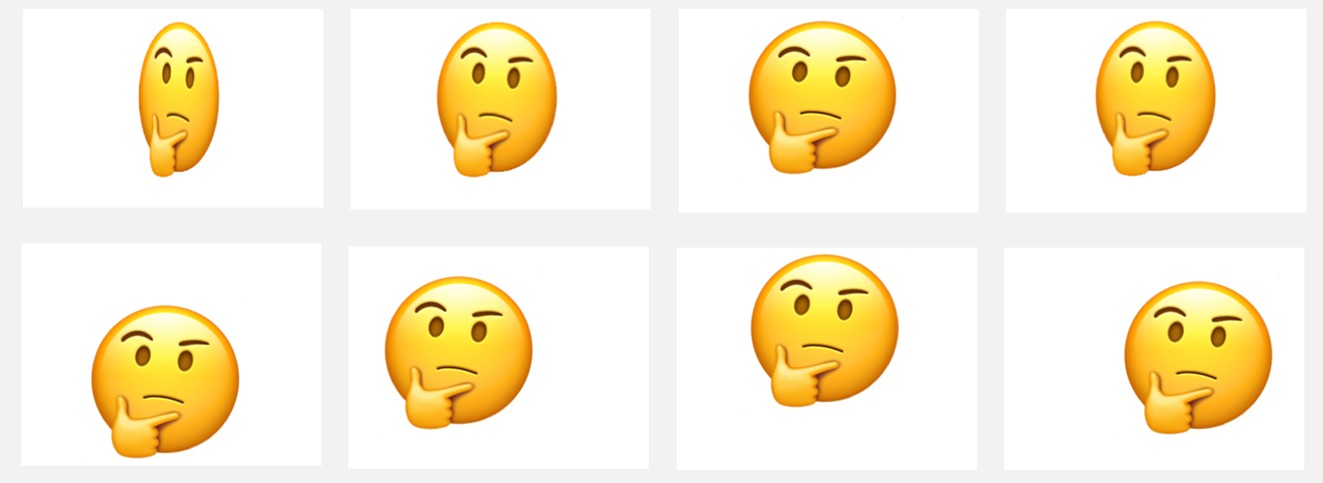}\\
  \bottomrule
  \end{tabular}
\end{table*}

\subsubsection{Theme 2: Users' Authoring Process---How They Select and Interpret Multi-modal Elements}\label{authoring_process}
This theme elaborates on participants' process of creating multi-modal emoticons. Namely, we summarize patterns from participants' rich explanations concerning their preferred orders and strategies of making selections, as well as their (common and differentiated) interpretations of the multi-modal elements.
To the best of our knowledge, no prior knowledge has been gathered regarding how users value and prioritize the different types of components when constructing a multi-modal emoticon. To probe a better understanding of this, we designed the \name{} interface to be open-ended: users could make selections across the three types of components in any order. 

\textbf{Which Element to Select First (and Why):} According to the users' reports (from both the interview and questionnaire data), a recurrent pattern of their selection order can be seen: in many cases, users would select an emoticon first, and then an animation a vibration (although meaningful exceptions were also reported). Detailed accounts have been provided by them. 
First, as agreed by the participants, emoticons often served as the base to set the tone for their multi-modal expression. As P5 stated, \qt{emoticons decide the theme of the emotion to express}, or, as P18 put, \qt{emojis are the major carrier of what I want to express.} In contrast, animations and vibrations were often used as nuanced modifications or further enhancement to the base meaning afforded by the emoticons. As they reasoned, this could be because emoticons are more direct and explicit, while animations and vibrations are more open-to-interpretation: \eg, \qt{when texting, I already associated my feeling with some emoji} (P2) but \qt{when selecting from vibrations, I do not have a particular one in my mind} (P20). Compared with animations, vibrations are even less concrete, since they are \qt{more subtle} (P19) and \qt{not so visual} (P20).

However, according to the participants, sometimes, animations could have the first-place contribution to the expressed meaning. For instance, P9 gave an example that he first decided to use the animation \textit{moving left and right} which felt like \qt{nudging him [P10] to wake him up} to ask for P10's response. Then he chose the emoticon \wsicon{images/emoji_sunglasses} only because it was fun. A few similar examples were also told by other participants.
In addition, P3 reported a different pattern: oftentimes, her second step is selecting a vibration, before looking for an animation. An extra criterion mentioned by several participants is that they sometimes tended to select animations and vibrations that could better match each other's temporal properties: such as rhythm or periodic duration. \eg, P15: \qt{I would make the vibration match the motion. [P16 added:] it's like dubbing a movie.} 

\textbf{Vibrations were Associated with Everyday Sounds, while Animations were Associated with Body Language:} Another noticeable pattern that emerged from the users' descriptions is how they used associations to explain their decision-making on selecting vibrations and animations. Namely, when reasoning about why they chose certain vibrations, they often associated their choices with certain everyday sounds; whereas, when explaining their choices for animations, they often used associations with body language. 
For instance, P3 explained that when constructing a multi-modal emoticon to convey greetings to P4, she selected a vibration that felt like the sound of \qt{knocking on the door.} P14 used a vibration to express a strong sad feeling because it felt like an \qt{electrical noise.} As another example, P15 explained that she chose a vibration that felt like the \qt{tick-tock} sound of a clock, to match the slow circular animation she selected. P9 also mentioned an example in which his selected vibration reminded him of the funny background music both he and P10 were familiar with. 
Other examples also included associations between vibrations and the sounds of snores, heartbeats, or human voices. 

Besides these, rich examples were gathered about how the animations were interpreted as body language. For example, P19 explained that she combined an animation of swinging left and right with a not-so-happy face \wsicon{images/emoji_expressionless} to express disagreement because it felt like \qt{shaking head left and right, disagreeing.} P17 added an animation of slight and quick shaking to the emoji \textit{grinning face with sweat} \wsicon{images/emoji_grinning_sweat} to express the meaning of embarrassment, since the motion resembled the \qt{face twitching when someone is embarrassed.} As both P17 and P18 agreed, the animations could make the expressions \qt{specific and concrete [...] more like adding body movements.} P9 similarly confirmed that when looking for animations, he did take \qt{body languages} into considerations, either consciously or subconsciously. 

\textbf{Diverse Individual Interpretations:} In addition to the common pattern of associating vibrations and animations respectively with sounds and bodily movements, individual differences also widely existed in the participants' interpretations. For instance, while P4 mentioned that she preferred to choose longer, more intense vibrations which felt richer and more interesting to her,  P20 thought that \qt{long vibrations feel angry.} For another example, while P12 considered that bigger motions could be used to represent happier feelings, P19 was concerned that bigger motions sometimes could appear too exaggerating and less honest. As P5 pointed out, such differences: \qt{could have something to do with [a user's] personality, as well as the topic of the conversation.} These individual differences in users' decision-making again, confirmed that it is meaningful to ensure a certain amount of open-endedness in design and allow users to freely combine various multi-modal emoticons based on their personal and contextual interpretations (\textbf{D1}).

\subsubsection{Theme 3: Users' Opinions on the Recommendation Approach}
This theme summarizes users' experiences and comments on the recommendation approach of the \name{} system, including the two-fold recommendation principles and the non-aggressive recommendation strategy.

\textbf{Why the Two-fold Principles were Meaningful:} The \name{} recommendation algorithm follows two-fold principles. One principle is to prioritize the multi-modal options that were more emotionally relevant (in terms of valence and arousal levels) to the user's selected element. 
The participants considered this principle to be intuitive and oftentimes meaningful. As confirmed by P15, \qt{it's pretty accurate, and it's similar to what I thought [...] for active and happy emojis, I tend to add faster and bigger motions.} P8 also thought that \qt{stronger emotions match stronger vibrations, weaker emotions match weaker vibrations.} They considered this principle to be helpful especially when users did not want to spend much time browsing the options. As stated by P19: \qt{some users are lazy, like me. I would use the recommendation. The options on the first screen are already sufficient} 

The other principle is to prioritize the elements that have been more frequently combined with the current section by the user in the past. The participants perceived this to be practical and efficient. For instance, to P17, \qt{historical frequency is pretty helpful, it's quite often that I reused some combinations.} And P3 felt the frequency-based principle \qt{pragmatic} and \qt{smart}: \qt{it's personalized, so it's about my own feelings.} In our design, the initial recommendations are based on emotional relevance, since no user data are gathered. Yet, over time, the frequency will shape the interface more and more, so that it would be adaptive to each user's personal preferences. This design decision has been explicitly confirmed by the participants. As P5 pinpointed, the emotional relevance could provide meaningful guidance to the users especially when \qt{in the beginning, no pattern [of use] has been formed}, and after a certain time, it is reasonable to \qt{meet each user's different needs, and be more personalized.} A very similar comment was also given by P16.

\textbf{Advantages of Non-aggressive Recommendation:} Following \textbf{D2}, the \name{} interface features a non-aggressive recommendation strategy that prioritizes the potentially relevant options, without automating users' decisions. In the interviews, the participants commented on this feature in comparison with some relatively more ``assertive'' recommendations they experienced from similar systems. For example, an emoji keyboard that predicts the ``best'' result to automate users' selection based on their text input. Overall, the participants had recognized two major advantages of the non-aggressive recommendation strategy over the more assertive ones in terms of multi-modal emoticons. 
First, the non-aggressive recommendation ensures users' personal expressions. As P18 put, \qt{if everyone is using this [recommended option] then why do I have to use it? I want to express my own emotions [...] I won't send the same one each time. I could try something new.} By prioritizing relevant options, users could be facilitated while still preserving sufficient variations for creating their own expressions. 
Second, an algorithm might not be successful all the time, due to its limited access to the whole conversation context. As P18 argued, a recommendation \qt{might not always be precise in predicting what you really want to express.} As telling examples for this, in \autoref{openended}, the participants reported a series of cases about how they combined multi-modal elements creatively to express meanings that are quite subtle, or different from the original meaning of the selected emojis. In this sense, a non-aggressive recommendation strategy could avoid giving arbitrary or irrelevant suggestions to users, while leaving space for users to make the modifications. These experiences thereby contextually verify our \textbf{D2}.

\subsubsection{Theme 4: Envisaged Opportunities---Multi-modal Emoticons Beyond Mobile Messaging}
This theme summarizes participants' envisions about desirable future scenarios or new application domains that multi-modal emoticons (emoticons augmented by animation effects and vibrotactile patterns) could be meaningfully leveraged, in addition to the mobile context. 

Interestingly, a number of their envisages applications went beyond human-human communication. For instance, P11 envisaged that in an AI-enhanced smart home system, a voice agent (\eg, Cortana by Microsoft, or Alexa by Amazon) could also utilize multi-modal emoticons to communicate with human users: \qt{apart from the voice interactions, if it could send me emojis with animation and vibrations [\eg, on mobile devices], it would feel different, it would feel more human, and richer information could be communicated..} P3 also envisaged a smart home context, in which the Internet-of-Things products could more expressively interact with users via multi-modal emoticons, and she thought a smartwatch could be a meaningful medium to receive vibrotactile signals. P4 from the same pair talked about a public setting: \qt{In some shopping malls, there are now AI robots for searching foods and places. If these robots could display emoticons with animations and vibrations, they would be more attractive to younger users I guess.} 

Moreover, P15 suggested using a multi-modal emoticon to summarize people's attitudes from internet reviews, \eg, of a restaurant: \qt{instead of reading all the reviews, you could then first feel the general reaction of people who went to eat, from an emoji with animation and vibration.} P7 and P16 similarly envisaged the usage of multi-modal emoticons in combination with the ``barrage captions'' in live-streaming, or live videos (a type of dynamic captions that are created by viewers in real-time and moving across the video to enable momentary communications between the live-streamer and viewers or among viewers). As P16 states, \qt{current live barrages are mainly texts, if dynamic emoticons and vibrations are added, it might be more fun and interactive.} In addition, both P5 and P12 envisaged that future authoring tools of multi-modal emoticons might consider supporting users to tap the touchscreen to define new vibrotactile patterns, or drawing a trajectory to create a new animation to further unleash users' creativity.

\section{Discussion}
This paper focuses on multi-modal emoticons: a trend burgeoning in recent social applications which enhances traditional static emoticons with animations or vibrotactile feedback. Multi-modal emoticons are increasingly introduced in current social applications. However, little empirical knowledge has been accumulated concerning how people create, share and experience multi-modal emoticons in daily communications. 
Addressing this opportunity, we have designed and evaluated the \name{} system to gather empirical understandings from real-world usage and experience, and surface design implications for supporting users' creativity and communication of multi-modal emoticons.

As shown in our field deployment, \name{} enabled a user-driven exploration. In various scenarios, the participants creatively combined stickers, animations, and vibrations to author multi-modal emoticons on the fly, which helped their spontaneous affective communication in the unfolding conversation. 
As indicated by the results, using multi-modal emoticons could make online chatting more fun, engaging, and expressive, increase their mental closeness, and potentially help them to communicate more specific, or richer emotional feelings. These experiences have been further supported and concertized by the participants' detailed descriptions in the qualitative findings. 
Beyond presenting a novel design case and contextually confirming the benefits of multi-modal emoticons, the rich empirical data also enabled us to have an in-depth analysis of the participants' creation and usage of multi-modal emoticons. 
We discuss the extracted underlying patterns and implications for future research and design below.

\textbf{Implication 1: Purposefully leveraging the extra affordance of multi-modal emoticons.} The variety of examples collected in the findings illustrate that by freely combining multi-modal elements, users were not only able to enhance or augment the expression of static emoticons, but also create new meanings that are additional to, or different from the intended meaning of static emoticons. This suggests that multi-modal emoticons can offer extra communication affordance to existing emoticon systems, implying new research and design opportunities. 

Studies have been conducted to explore the idea of \textit{emoji affordance} on how users could build upon the visual or contextual properties of an emoji to enable richer meanings \cite{Wiseman2018_repurposingemoji, Khandekar2019_emojifirst, alvina2019mojiboard}.
However, little has been done to understand and leverage the communication affordance of multi-modal emoticons.
While serving as an early exploration, our work has surfaced that the multi-modal elements like animations and vibrotactile signals can facilitate users' communication beyond simply as subsidiary feedback to specific emoticons (which is how they have often been designed in current messaging apps). Instead, just like new meanings could emerge from the combination of multiple emoticons \cite{Khandekar2019_emojifirst}, our results suggest that new expressions could be created by combining static emoticons with different animations and vibrations. We thereby argue that future designs could purposefully help users leverage this new affordance to meaningfully expand their nonverbal communication vocabulary. To do so, a user-led approach is needed: researchers and designers could continually curate and analyze examples of users' creative expressions via multi-modal emoticons (\eg, \autoref{openended}), and intentionally facilitate similar use cases in design.

\textbf{Implication 2: Designing for immersive and empathetic experience in online nonverbal communication.} 
As uncovered by our empirical data, another promising opportunity of multi-modal emoticons is that they could create a certain atmosphere for a chat session, or make their responses feel more empathetic and attentive to each other. These new experiences in nonverbal communication were not supported by static emoticons. 
For instance, P15 always used a specific multi-modal emoticon to set the mood for joking, and P9 used multi-modal emoticons to render friendly, and relaxing vibes. Such immersive, and \qt{contagious} ambience was, to a large extent, enabled by the animation effect which was continuously rendered as long as the emoticon is displayed on the chat screen. 
The future design could build upon this experiential quality and make online communication even more immersive: \eg, when a new multi-modal emoticon is sent, the ambient visual elements on the chat screen (\eg, chat bubbles or background) will react to it and adjust to the same atmosphere (\eg, a happy multi-modal emoticon will make the ambient elements look brighter), until the next emoticon shapes the ambience again. 

As reasoned by the participants, the attentive, empathetic responses were enabled partially because they could create small variations of animations or vibrations in the multi-modal emoticons, instead of sending receptive or preset emoticons. Whereas, responding to others with the repeated static emoticon might feel in-polite and in-attentive. The empathetic experience was reported to be further enhanced when the two conversation partners used similar animation or vibration to express similar feelings, which made them feel more \qt{connected} and \qt{in sync}. Inspired by this, the future design could explore enhancing such empathetic experience, \eg, by creating a resonance, or echo effect on the screen, when two users used similar multi-modal elements in chat.

\textbf{Implication 3: Designing new animations and vibrotactile patterns based on body languages and everyday sounds.} 
The great creativity exhibited by our participants in authoring multi-modal emoticons was supported by the building blocks: the sets of animations and vibrations prepared upon the open libraries of Kineticons \cite{Harrison2011} and VibViz \cite{Seifi2015}. Despite being comprehensive and well-designed, these libraries were intended to support a wide range of design purposes in graphical or haptic interfaces, rather than multi-modal emoticons. For this reason, as mentioned in \autoref{sec:elements}, not the whole libraries could be meaningfully used in \name{}. Hence, a meaningful question in this emerging domain is: how we can design new building blocks, such as libraries of animation effects or vibrotactile patterns, to better nourish the creativity of both designers and everyday users of multi-modal emoticons.

Our findings from \autoref{authoring_process} could shed light on this question.
Participants often associated their choice for vibrations with certain everyday sounds (\eg, knocking on the door, snoring, or tick-tock). Moreover, their choice for animations was often associated with bodily movements or body language (\eg, nodding, pushing, face-twitching). When there was a match between their communication intention and the association of the element, they would choose that element (\eg, P19 used a \qt{nodding animation} to express agreement or hospitality; P3 used a \qt{snore vibration} to express going to sleep). This interesting pattern suggests that new animations and vibrotactile signals could be designed to enable users' richer associations with the two categories. 
For instance, future work could design new vibration elements based on a typical set of everyday sound effects. Or, designers could refer to a taxonomy of (bodily) nonverbal signals \cite{vinciarelli2009social} to design new animations that could communicate more intentions of users (\eg, resembling certain eye behaviors or arm gestures, in addition to the head poses or full-body motions that they often mentioned). 

\textbf{Implication 4: balancing the design trade-off of automation versus autonomy.} 
To support user-authoring multi-modal emoticons, \name{} features a non-aggressive recommendation approach, which prioritizes relevant options for users without limiting their free experimentation. As appreciated by the participants, such an approach leaves enough agency for them to create personalized expressions and avoids giving arbitrary and irrelevant suggestions due to misinterpreting the user's intention or the context. 
However, we do recognize that a design tension exists, concerning the trade-offs between automation and autonomy. For instance, a number of prior works explored fully automating users' entry of emoticons based on facial expressions \cite{Liu2018_reactionbot,Ali2017_face2emoji} or speech \cite{Hu2019_speechtoemoji}. 

While automation often means low effort and high efficiency, our study revealed that in certain social communication scenarios, users value the autonomy and richness of self-expression more than efficiency. As reported earlier, while considering the recommendation principles, the users enjoyed being explorative and exploring different combinations. Therefore, an important consideration for future work is to better understand users' different needs for autonomy and automation in various social communication scenarios and carefully balance the two in design.

\textbf{Limitations.} Our work is not without limitations. First, \name{} included the same 50 (frequently used) emojis for each user to ensure that they had the same building blocks to begin with. Although abundant examples of their creative usage have been found, users could have more freedom if they could import new stickers to the system. A future field study can be conducted to enable each pair to co-design their libraries, in the form of the co-customization \cite{Griggio2021} for multi-modal emoticons. 
Second, participants in this field study were aged from 19 to 35; hence, how different age groups will use multi-modal emoticons and to what extent the findings might be applicable beyond the current user segment remain to be answered. 
It would be interesting in the future to involve older age groups or even elderly users, to understand their experiences. Additionally, the \name{} user interface might be further polished: its size could be made a bit smaller to display more historical chats above, and a search bar could be added to enable element selection by keyword. 
Lastly, \name{} currently employs a recommendation algorithm that only considers the usage frequency and patterns of one single user (\ie, the benefiting user). With more data collected, advanced recommendation techniques, such as collaborative filtering \cite{koren2015advances}, can be achieved by mining common patterns across all the users and suggesting frequently-used multi-modal emoticon combinations of the crowd that is similar to the benefiting user. 

\section{Conclusion}
In this study, we set out to empirically understand multi-modal emoticons: an emerging phenomenon in current messaging applications which enhance static emoticons with multi-modal effects such as animations and vibrations. 
In doing so, we built \name{}, a user-authoring multi-modal emoticon system for mobile messaging. The design of \name{} extends beyond the current systems and allows us to extract new understandings about how users create, share, and experience multi-modal emoticons in daily communications, and how we could better support them by design. 
We evaluated \name{} with 20 participants in naturalistic settings over the period of four weeks. Rich findings were gathered to contextualize the creative usage and meaningful scenarios of multi-modal emoticons, as well as the users' detailed process of authoring. Based on the rich data, we generalize relevant design implications to inform future work in better supporting users' creation and communication of multi-modal emoticons.

\begin{acks}
We thank Dr. Wei Li from the Human-Machine Interaction Lab, Huawei Canada, for his support and contribution to this project since its very early stage. We thank all our participants for their time and valuable inputs. We would also like to thank our reviewers whose generous and insightful comments have led to a great improvement of this paper. 
\end{acks}

\bibliographystyle{ACM-Reference-Format}
\bibliography{reference.bib}

\end{document}